\RequirePackage{fixltx2e}
\documentclass[reprint, amssymb, amsmath, amsfonts, a4paper, aps, floatfix, superscriptaddress, showpacs, twoside]{revtex4-1}
\usepackage{tabularx, xspace, rotating, units, enumerate, textcomp, icomma, hyperref}

\usepackage[T1]{fontenc}

\newcommand{\Epi}{\affiliation{Department of Epileptology, University of Bonn, Sigmund-Freud-Stra{\ss}e~25, 53105~Bonn, Germany}}
\newcommand{\HISKP}{\affiliation{Helmholtz Institute for Radiation and Nuclear Physics, University of Bonn, Nussallee~14--16, 53115~Bonn, Germany}}
\newcommand{\IZKS}{\affiliation {Interdisciplinary Center for Complex Systems, University of Bonn, Br\"uhler Stra\ss{}e~7, 53175~Bonn, Germany}}
\newcommand{\ICBM}{\affiliation {Theoretical Physics/Complex Systems, ICBM, Carl von Ossietzky University of Oldenburg, \\Carl-von-Ossietzky-Stra\ss{}e~9--11, Box~2503, 26111~Oldenburg, Germany}}
\newcommand{\RNS}{\affiliation{Research Center Neurosensory Science, Carl von Ossietzky University of Oldenburg,\\ Carl-von-Ossietzky-Stra\ss{}e~9--11, 26111~Oldenburg, Germany}}

\newcommand{\abs}[1]{\mathopen{}\mathclose\bgroup\left\lvert#1\aftergroup\egroup\right\rvert}
\newcommand{\norm}[1]{\mathopen{}\mathclose\bgroup\left\|#1\aftergroup\egroup\right\|}
\newcommand{\kl}[1]{\mathopen{}\mathclose\bgroup\left(#1\aftergroup\egroup\right)}
\newcommand{\klg}[1]{\mathopen{}\mathclose\bgroup\left\{#1\aftergroup\egroup\right\}}
\newcommand{\kle}[1]{\mathopen{}\mathclose\bgroup\left[#1\aftergroup\egroup\right]}
\newcommand{\kls}[1]{\mathopen{}\mathclose\bgroup\left\langle#1\aftergroup\egroup\right\rangle}

\newcommand{\defi}{\mathrel{\mathop:}=}

\newcommand{\AtoB}{A\textrightarrow B\xspace}
\newcommand{\AtoC}{A\textrightarrow C\xspace}
\newcommand{\BtoA}{B\textrightarrow A\xspace}
\newcommand{\BtoC}{B\textrightarrow C\xspace}
\newcommand{\CtoA}{C\textrightarrow A\xspace}
\newcommand{\CtoB}{C\textrightarrow B\xspace}

\newcommand{\Atilde}{\smash{\(\mathfrak{A}\)}\xspace}
\newcommand{\Btilde}{\smash{\(\mathfrak{B}\)}\xspace}
\newcommand{\Ctilde}{\smash{\(\mathfrak{C}\)}\xspace}

\begin{document}

\title{Self-induced switchings between multiple space--time patterns on complex networks of excitable units}

\author{Gerrit Ansmann}
\Epi \HISKP \IZKS

\author{Klaus Lehnertz}
\Epi \HISKP \IZKS

\author{Ulrike Feudel}
\ICBM \RNS

\begin{abstract}
We report on self-induced switchings between multiple distinct space--time patterns in the dynamics of a spatially extended excitable system.
These switchings between low-amplitude oscillations, nonlinear waves, and extreme events strongly resemble a random process, although the system is deterministic.
We show that a chaotic saddle---which contains all the patterns as well as channel-like structures that mediate the transitions between them---is the backbone of such a pattern switching dynamics.
Our analyses indicate that essential ingredients for the observed phenomena are that the system behaves like an inhomogeneous oscillatory medium that is capable of self-generating spatially localized excitations and that is dominated by short-range connections but also features long-range connections.
With our findings, we present an alternative to the well-known ways to obtain self-induced pattern switching, namely noise-induced attractor hopping, heteroclinic orbits, and adaptation to an external signal.
This alternative way can be expected to improve our understanding of pattern switchings in spatially extended natural dynamical systems like the brain and the heart.
\end{abstract}

\pacs{05.45.Xt, 05.90.+m, 89.75.Kd, 89.75.Hc }

\maketitle

\section{Introduction}
Pattern formation processes in spatially extended nonlinear systems have been a long-standing topic of extensive theoretical and experimental research \cite{Cross1993, Hoyle2006}.
Patterns that are fixed in space can be either time-independent like Turing patterns in biology~\cite{Turing1952}, chemistry~\cite{Agladze1992, Perraud1992, Ruediger2003} and ecology~\cite{Baurmann2004, Baurmann2007} or time-dependent like convection rolls in hydrodynamics~\cite{Berge1983}.
Another prominent example of pattern formation concerns nonlinear waves in neuroscience \cite{*[][{ and references therein}] Ermentrout2010} and in the Belousov--Zhabotinsky reaction \cite{Vanag2001, Mikhailov2006}.
More recently the formation of localized structures in spatially extended systems has been studied using paradigmatic models like the complex Ginzburg--Landau equation~\cite{Uecker2014}, the Swift--Hohenberg equation \cite{Gelens2011}, or networks of bistable elements~\cite{Kouvaris2013}.
Examples of patterns in nature are vegetation patterns in semiarid areas \cite{Hardenberg2001, Gilad2004}, patterns of mussels on tidal flats \cite{Koppel2008, Liu2014}, waves, bursts and other spontaneous activities in the brain \cite{Engel2001, Mazzoni2007, Ringach2009, Sato2012} and the heart~\cite{*[][{ and references therein}] Bub2005}, or circulation patterns in the atmosphere \cite{Franzke2011, Franzke2013} and in the ocean \cite{Schmeits2001, Pierini2009}.

While most of these patterns correspond to single attractors to which all initial conditions converge, there are also several systems in nature that possess coexisting attractors corresponding to different stable space--time patterns.
Which of these patterns is finally realized depends crucially on the initial conditions.
As an example we mention the bistable vegetation model describing stripes and spots of vegetation in a desert which can coexist with the state of bare soil \cite{Gilad2007}.

Moreover, natural systems often exhibit a switching between different patterns, which can occur due to different reasons.
Here we do not refer to pattern changes that occur because one pattern loses stability and another pattern gains it due to a change of an external parameter.
A vast literature on the mathematical treatment of such bifurcations exists~\cite{Cross1993} and they have also been studied in natural systems like ecosystems or the climate system, where they are often called \textit{regime shifts} or \textit{tipping points,} respectively \cite{Dakos2008, Lenton2008, Scheffer2012}.
We also do not refer to cases, in which certain parts of a system exhibit a switching of patterns, but the system regarded as a whole remains in a state where different local patterns coexist, such as spatiotemporal intermittency~\cite{Chate1994}, spatiotemporal chaos~\cite{Gollub1995}, or moving chimera states~\cite{Omelchenko2010}.

By contrast, several natural systems exhibit intermittent pattern switchings that affect the whole system and cannot be attributed to a parameter change.
The main characteristic of these switchings is the irregularity in their timing.
They have been observed amongst others in Rayleigh--B\'enard convection~\cite{Berge1983}, the nonlinear Schr\"odinger equation~\cite{Anderson2014}, in random networks of spiking neurons \cite{Kriener2014, Rothkegel2011}, and in the complex Ginzburg--Landau equation \cite{Popov2007, Hai2010} and a modification thereof~\cite{Haugland2015}.
If a certain pattern in such a system is sufficiently rare, short-lived, and can be related to a strong impact, it may be regarded as an extreme event \cite{Albeverio2006, Ghil2011}.

There are three known mechanisms for such switchings:
\begin{itemize}
\item
In a multistable system, i.e., a system in which several attractors coexist, small noise can kick the dynamics away from an attractor to the boundary of its basin, from which the dynamics will either return to the previous one or move to another attractor and thus the pattern will switch.
This noise-induced switching process, which is also called \textit{attractor hopping}~\cite{Kraut2003} or \textit{chaotic itinerancy}~\cite{Kaneko2003, Masoller2002}, has been found in spatially extended systems governed, e.g., by the complex Ginzburg--Landau equation~\cite{Coullet2004}, but also in neuron networks~\cite{Marro2007}, genetic networks~\cite{Labavic2014}, networks of phase oscillators~\cite{DHuys2014} or chemical systems~\cite{Mertens1994}, where this phenomenon is denoted as \textit{intermittent chemical turbulence.}
A common property of all those noise-induced switchings is the exponential distribution of the residence times in the vicinity of a specific attractor \cite{Kramers1940, Haenggi1990}, i.e., of the pattern durations.
\item
If the system possesses a heteroclinic orbit, a typical trajectory traverses the neighborhood of all saddle states connected by this orbit \cite{Afraimovich2004, Ashwin2007} and thus the dynamics switches between the corresponding patterns.
While the sequence of saddle states is always the same, the time spent close to each of them increases as time goes by, being infinite on the heteroclinic orbit itself.
Among the best-studied systems of this kind are neuronal systems, where such a sequential activity has been successfully modeled \cite{Komarov2013, Levanova2013} and related to experimental observations~\cite{Rabinovich2008}.
\item
As a third mechanism, we mention the adaptation to an external signal, which is extensively discussed with respect to perception processes in the brain~\cite{Kriener2014}.
Here, the basic assumption is that either the adaptation to a time-dependent stimulus~\cite{Aucouturier2008} or the competition between different activity regimes for a fixed stimulus~\cite{Rankin2014} gives rise to an intermittent switching between different neuronal patters.
\end{itemize}

In this paper we show that the collective dynamics of complex networks of excitable units can exhibit self-induced intermittent switchings between multiple space--time patterns that are neither related to a heteroclinic orbit nor are they induced by noise or an external signal.
The space--time patterns we observe are low-amplitude oscillations, nonlinear waves, and short-lived extreme events.
We show that the switching dynamics is facilitated by channel-like structures on a chaotic saddle that contains all the patterns.
We attribute this pattern switching to two properties of the system:
(i)~the capability of assuming a state in which it resembles an inhomogeneous, oscillatory medium and in which it can self-generate spatially localized excitations and
(ii)~the dominance of short-range connections and a certain amount of long-range connections.

In Sec.~\ref{phenomenological} we explain the employed model and phenomenologically describe the switching dynamics.
We investigate in Sec.~\ref{parameters} how control parameters influence the dynamics, and in Sec.~\ref{statistics}, we analyze recurrences and lifetimes of patterns.
In Sec.~\ref{channels}, we present the chaotic saddle and investigate its internal structure.
In Sec.~\ref{discussion}, we discuss our findings, before we draw our conclusions.

\section{Model and first observations}\label{phenomenological}

As a paradigmatic model for excitability, we consider diffusively coupled FitzHugh--Nagumo oscillators \cite{FitzHugh1961, Nagumo1962}, which we denote as \emph{units.}
Unless mentioned otherwise, our system consists of \(n=10000\) such units, each of which we associate with one node of an unweighted, undirected small-world network~\cite{Watts1998} based on an \(100 \times 100\) lattice with cyclic boundary conditions (torus).
We connect each unit to each other unit within a two-dimensional \emph{sphere of local influence} and choose the radius such that the sphere contains \(m=60\) units.
Each connection was rewired with a probability of \(p=0.2\), i.e., it was removed and replaced by a connection between two randomly chosen units (avoiding self-connections and duplicate connections).
The dynamics of unit~\(i\) is governed by the following differential equations:
	\begin{align}
		\dot{x}_i & = x_i (a-x_i) (x_i-1) - y_i + \frac{k}{m} \sum\limits_{j=1}^{n} A_{ij} (x_j - x_i), \notag\\
		\dot{y}_i & = b_i x_i - c y_i.
		\label{eq:systems}
	\end{align}
\(x_i\) is the excitatory and \(y_i\) the inhibitory variable.
The constant internal parameters are \(a=-0.0276\) and \(c=0.02\); \(b_i\) is drawn from the uniform distribution on \(\kle{0.01-\tfrac{1}{2} \Delta b, 0.01+\tfrac{1}{2} \Delta b}\) for each~\(i\), with \(\Delta b = 0.008\).
\(A \in \klg{0,1}^{n \times n}\) is the adjacency matrix of the small-world network, and the coupling strength is the quotient of \(k=0.128\) and \(m\).

The system was realized and integrated with the software package Conedy~\cite{Rothkegel2012}, using an adaptive Runge--Kutta--Fehlberg procedure.
We exemplarily checked that the following observations are neither specific to the chosen integrator nor the numerical accuracy.
Initial conditions were chosen independently for each unit from a normal distribution with mean \(0.34\) and standard deviation \(0.60\) for \(x_i\) and mean \(0.15\) and standard deviation \(0.1\) for \(y_i\).
We are confident that this choice does not affect the range of observable dynamical behaviors, as the initial conditions covered about half of that part of the phase space that is relevant to our dynamics.

If a unit with any of the possible~\(b_i\) were uncoupled, it would exhibit periodic oscillations, and if the parameter~\(b\) were homogeneous, all units would exhibit this behavior even if coupled, being in complete synchrony.
Coupling and inhomogeneity together, however, have an inhibitory influence on every individual unit and cause it to exhibit, in general, an excitable behavior, which we will describe in detail in the following.

\begin{figure*}
	\includegraphics{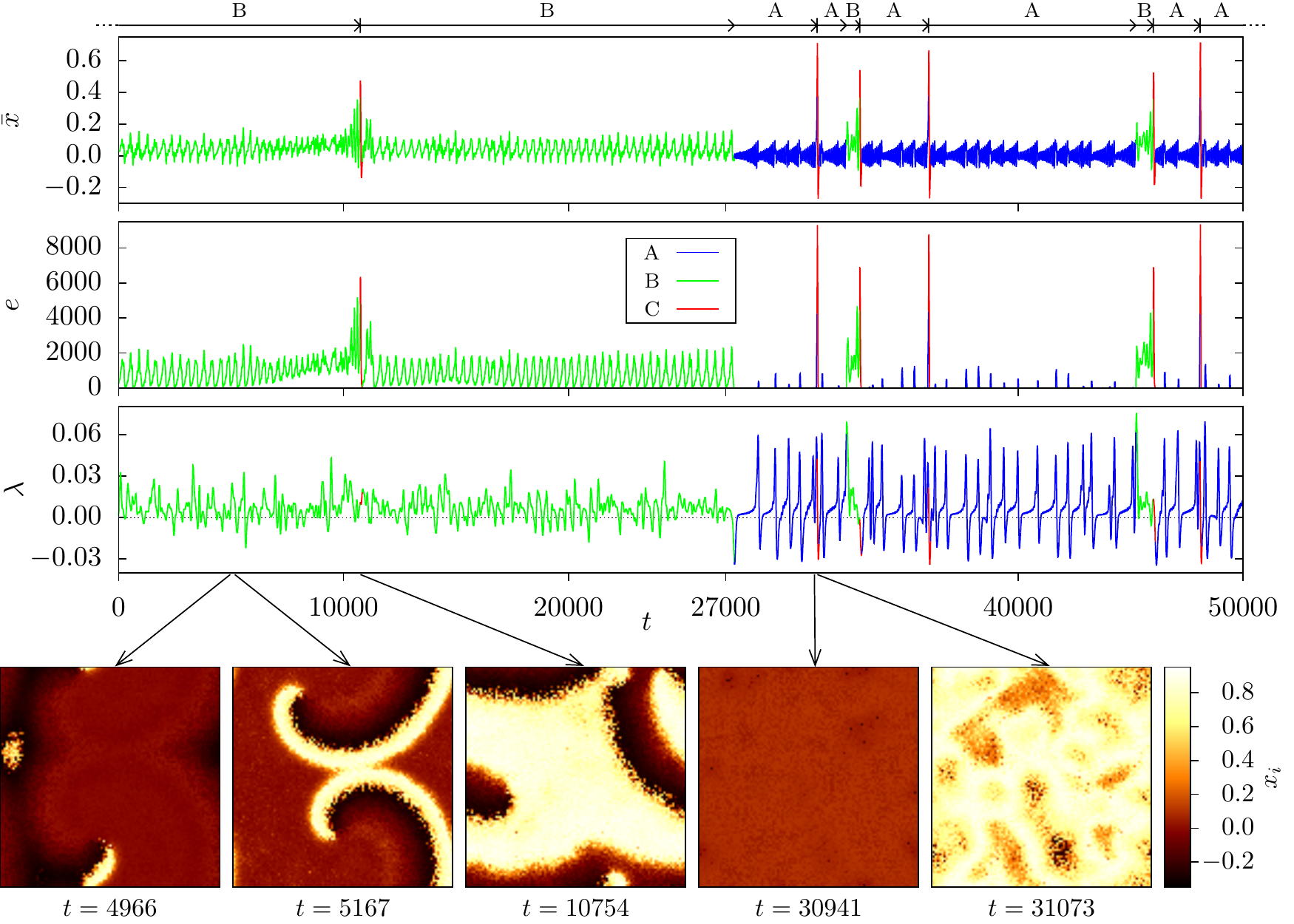}
	\caption{
	(First to third row): Exemplary temporal evolutions of~\(\bar{x}\), of the number~\(e\) of units with \(x_i>0.4\) (``excited units'') and of an estimate~\(\lambda\) of the largest local Lyapunov exponent (temporal evolution smoothed with a Gaussian kernel with a width of \(30\) to improve readability).
	The line colors indicate the patterns as automatically classified (blue:~A, low-amplitude oscillations; green:~B, waves; red:~C, extreme events).
	These patterns are also indicated at the very top with pattern~C being indicated with a vertical line.
	(Bottom) Snapshots of the spatial distribution of~\(x_i\kl{t}\) at times corresponding to selected local minima and maxima of~\(\bar{x}\) (from left to right): adjacent minimum and maximum around \(t=5000\), the maximum during the event around \(t=10000\), the minimum before the event around \(t=30000\) and the maximum during that event.
	Units are represented by pixels, which are arranged according to the lattice underlying the small-world network and whose color encodes the value of the respective~\(x_i\).
	(See the \href{http://arxiv.org/src/1602.02177v2/anc/animation.avi}{the ancilliary file} for an animation for the entire shown time period.)
	\label{fig:zeitreihen}
}
\end{figure*}

In the first and second row of Fig.~\ref{fig:zeitreihen}, we show a typical temporal evolution of the average value of the first dynamical variable, \(\bar{x} = \tfrac{1}{n} \sum_{i=1}^n x_i\) and the number~\(e\) of excited units (i.e., units with \(x_i\) exceeding a certain threshold) along with snapshots of the individual units'~\(x_i\) (bottom row; see also \href{http://arxiv.org/src/1602.02177v2/anc/animation.avi}{animation in the ancilliary files}).
We observe the system's dynamics to switch between three different \textit{space--time patterns,} which we detail in the following based on this exemplary trajectory:

For \(0 < t\lessapprox 10000\), \textbf{waves} of excitation propagate over the torus.
Such waves are a common behavior for excitable media~\cite{Mikhailov2006}.
Wavefronts grow in spirals and are partially, but not fully destroyed upon collision with each other and due to the remaining excitation the waves do not die out.
This behavior roughly repeats about every 300 time units (which we refer to as a \textit{repetition} in the following) with the deviations from strict periodicity including a slow wandering of spiral centers.

At \(t \approx 10000\), small localized regions of units become excited simultaneously without a wavefront propagating over them.
The size of these regions quickly increases with each repetition, which finally culminates in an \textbf{extreme event} at \(t \approx 10754\) where almost all units become excited simultaneously and \(\bar{x}\)~exhibits an unusually large value.
This excitation subsequently dies out, leaving only a few excited units, from which wavefronts start to propagate once more, resulting in a behavior similar to that seen for \(0 < t\lessapprox 10000\).

\begin{figure*}
	\includegraphics{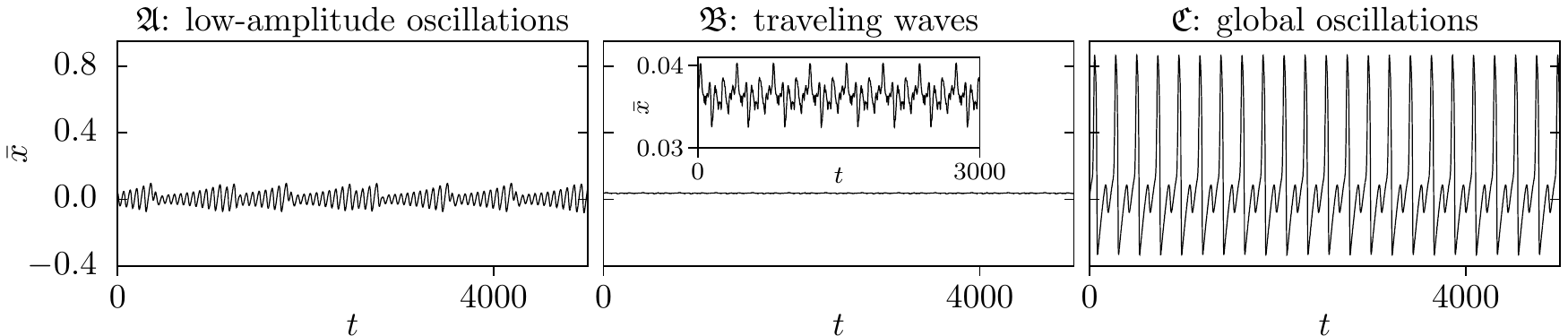}
	\caption{Exemplary temporal evolutions of \(\bar{x}\) for the attractors \Atilde, \Btilde and \Ctilde.}
	\label{fig:attraktoren}
\end{figure*}

Approaching \(t\approx 27000\), there is a decrease of the minimal number of excited units during each repetition; this decrease occurs after two wavefronts have almost canceled out each other.
At \(t\approx 27000\) the two wavefronts eventually fully cancel out each other upon collision, leaving no excited unit in the system.
Afterwards, for \(27000 \lessapprox t \lessapprox 31000\), all units perform roughly synchronous \textbf{low-amplitude oscillations} with a period length of about 70 time units.
The envelope of these oscillations exhibits an irregular sawtooth-like behavior with a small patch of units getting shortly excited at its local maxima but without starting a propagating excitation~\footnote{This behavior resembles the low-amplitude oscillations and proto-events observed in systems that are comparable except for having a completely coupled topology~\cite{Ansmann2013}.}.
At \(t \approx 31000\), this excitation extends to more units than before, and subsequently almost all units become excited simultaneously, constituting another extreme event.
Afterwards, all units become refractory, no excited units remain in the system, and the system returns to low-amplitude oscillations.
At \(t \approx 32000\), a small patch of units become excited during the low-amplitude oscillations in the aforementioned manner.
Instead of ceasing or spreading quickly, these excitations propagate in a wave-like manner leading to a behavior comparable to that for \(0 < t\lessapprox 10000\).
From then on, we only observe space--time patterns and switchings between them as already described.

To analyze this switching process, we formulate criteria that allow to automatically distinguish the observed space--time patterns.
For this purpose, we employ parameters \(\theta_\text{E}\), \(\theta_\text{D}\), and \(\theta_\text{W}\), which are related to characteristic time scales of the patterns' dynamics:
\begin{description}
	\item [Pattern C -- \textit{extreme events}] A time interval beginning with \(\bar{x}\) rising above \(x_\text{thr}\) and with a length of exactly~\(\theta_\text{E}\).
	If another occurrence of pattern~C would begin less than a time span~\(\theta_\text{D}\) after the end of the last instance of this pattern, we regard both occurrences and the time in between as a single occurrence of pattern~C.
	\item [Pattern B -- \textit{waves}] A time interval longer than~\(\theta_\text{W}\) during which \(e \geq 1\) and no extreme event happens.
	\item [Pattern A -- \textit{low-amplitude oscillations}] A time interval during which the criteria for patterns C and~B are not met.
\end{description}
Note that by these definitions, two occurrences of one pattern need to be separated by an occurrence of another pattern.

In the following, we set \(x_\text{thr}=0.4\), which corresponds to the threshold of excitation of a single unit.
\(\theta_\text{E} = 90\) corresponds to the typical duration of a single extreme event (as identified by its excursion in phase space).
To capture possible sequences of extreme events in short succession as one occurrence of pattern~C, we set \(\theta_\text{D}=280\).
Setting \(\theta_\text{W}=560\) guarantees that the wave state exhibits at least one repetition (see above) as \(\theta_\text{W}\) is larger than the time a wavefront needs to travel around the torus and larger than the average observed repetition time.
Hereby, we avoid to identify as pattern~B short-lived excitations occurring frequently during low-amplitude oscillations as well as short-lived excitations at the beginning and after extreme events.
Our results were robust to small changes to \(x_\text{thr}\), \(\theta_\text{D}\) and~\(\theta_\text{W}\) and not affected at all by small changes to \(\theta_\text{E}\).

We indicate the patterns as automatically classified according to the above definitions with different line colors and labels atop of Fig.~\ref{fig:zeitreihen}.
For this and other examples, these comply with the results of visual inspection.
We refer to a behavior as exemplified in Fig.~\ref{fig:zeitreihen}, i.e., the system switching between all three patterns (A, B, and~C), as \textit{pattern switching.}

As a first indicator that the patterns also differ dynamically, we show an instantaneous estimate of the largest local Lyapunov exponent~\(\lambda\) in the third row of Fig.~\ref{fig:zeitreihen} (cf.~\cite{Grassberger1988b, Abarbanel1991}; see Appendix~\ref{lyap_local} for details).
We observe a characteristic difference in the temporal evolution of~\(\lambda\) between longer lasting occurrences of pattern A and~B:
While in both cases, \(\lambda\)~exhibits oscillations, these are slower, have a higher amplitude and a tangens-like shape for pattern~A~\footnote{The reason for this behavior could be the specific dynamics of the mean value~\(\bar{x}\) for our model:
During the time span in which the dynamics of the mean value is rather regular, i.e., oscillatory with increasing amplitude, the local Lyapunov exponent is close to~\(0\), while it develops a spike when the amplitude of this oscillation is abruptly reset to low values.}.
The largest time-averaged Lyapunov exponent of the dynamics is positive (0.0065, see Appendix~\ref{lyap_avg}), indicating a chaotic dynamics.

In order to shed light on possible mechanisms of pattern switching, we analyze characteristics of the dynamics as well as their dependencies on properties of the system using long trajectories (Sec.~\ref{parameters}: \(5 \cdot 10^4\) time units, discarding \(1.5 \cdot 10^5\) initial time units~\footnote{We chose the number of discarded time units this large due to the long-lasting transients that can be observed in our system (see Sec.~\ref{statistics}.)};
Secs. \ref{statistics} and \ref{channels}: \(2\cdot 10^6\) time units, discarding \(10^4\) initial time units).

If only one pattern occurred during the observation time, we consider the dynamics to have converged to an attractor.
We can distinguish three types of such attractors (see Fig.~\ref{fig:attraktoren}), which we refer to as \textit{low-amplitude oscillations}~(\Atilde), \textit{traveling waves}~(\Btilde) and \textit{global oscillations}~(\Ctilde).
Attractor~\Atilde has the same characteristics as pattern~A, which occurs within pattern switching, and can thus be considered its stable counterpart (like, e.g., stable and unstable fixed points).
An attractor of type~\Btilde corresponds to a single straight wavefront spanning the whole underlying torus.
There are at least eight coexisting attractors of this type, which differ in the orientation of the wave (one for each cardinal and intercardinal direction).
In contrast to the straight wavefronts of attractor~\Btilde, wavefronts occurring during pattern~B within pattern switching are spiral-shaped and have a smaller length.
Attractors of type~\Ctilde are periodic, global mixed-mode oscillations~\cite{Desroches2012}.
Each of their high-amplitude oscillations resembles a single event, as seen during pattern~C within pattern switching.
In contrast to attractor~\Ctilde, pattern~C usually consists of only one such event (or at most six).
Almost all trajectories identified with attractor~\Btilde and all trajectories identified with attractor~\Ctilde were periodic according to the test proposed in Ref.~\cite{Ansmann2015b}.
Moreover, all investigated instances of these attractors have a time-averaged largest Lyapunov exponent of zero (see Appendix~\ref{lyap_avg}).

\begin{figure*}
	\includegraphics[scale=0.5]{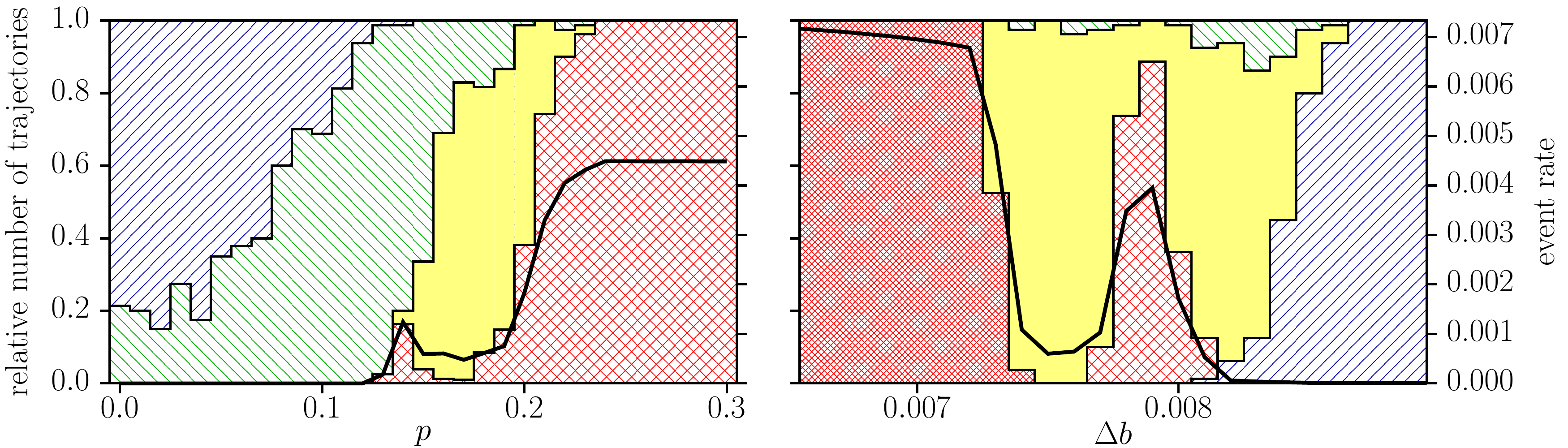}
	\caption{
Stacked histograms showing the relative number of trajectories exhibiting either pattern switching (yellow, solid) or convergence to one of the attractors \Atilde, \Btilde, or \Ctilde
(\Atilde, low-amplitude oscillations: blue, north-east diagonals;
\Btilde, traveling waves: green, south-east diagonals;
\Ctilde, global oscillations: red, cross-hatched).
Left: Influence of rewiring probability (\(p\), bin size: \(0.01\)) with a fixed parameter inhomogeneity (\(\Delta b = 0.008\));
Right: Influence of parameter inhomogeneity (\(\Delta b\), bin size: \(0.0001\)) with a fixed rewiring probability (\(p=0.2\)).
For each value of \(p\) and \(\Delta b\), the total number of trajectories was \(80\) (except for \(p<0.1\) and \(p>0.24\), where it was \(40\)).
For each of those trajectories, we employ a different realization of the coupling topology (corresponding to that~\(p\)) and of the inhomogeneity (corresponding to that~\(\Delta b\)).
	In the right panel, narrow cross-hatching indicates 1\textsuperscript{0}~global oscillations, and wide cross-hatching indicates 1\textsuperscript{1}~global oscillations~\cite{Note5}.
	The black thick line is the event rate (number of times at which~\(\bar{x}\) rises above \(x_\text{thr}\) divided by the observation time) averaged over all realizations.
	Almost all trajectories exhibiting pattern switching featured all three patterns A, B and~C.
	}
	\label{fig:p_dependency}
\end{figure*}

\section{Impact of the coupling topology and inhomogeneity}\label{parameters}

Based on previous observations \cite{Ansmann2013, Karnatak2014}, we conjecture that the coupling topology and inhomogeneities in the units might be important ingredients for pattern switching to occur.
In the following, we investigate the influence of these properties on the dynamics of the system.

\subsection{Coupling topology}\label{rewiring}

The topology of our network is determined by two properties:
the number of long-range connections, which is controlled by the rewiring probability~\(p\), and the maximum radius of short-range connections, which can be manipulated by resizing the sphere of local influence (via~\(m\)).

\emph{Amount of long-range connections}
--~Let us first discuss, how pattern switching or convergence to an attractor is affected by the amount of long-range connections (left panel of Fig.~\ref{fig:p_dependency}).
If the coupling topology features only short-range connections (\(p=0.0\)), the predominant dynamics is a convergence to the low-amplitude-oscillations attractor~(\Atilde).
If we add long-range connections, we observe the following sequence of regimes in parameter space, (as characterized by their predominant dynamics): traveling waves~(\Btilde), pattern switching, and finally global oscillations~(\Ctilde).
The latter manifests as a 1\textsuperscript{1}~mixed-mode oscillation~\footnote{	We here employ the \(H^L\) notation for mixed-mode oscillations with \(H\)~denoting the number of high-amplitude oscillations and \(L\)~denoting the number of low-amplitude oscillations in one period.
}
and lasts all the way to a random network (\(p=1.0\), not shown).

We explain the observed sequence of regimes as follows:
With an increasing inhomogeneity of the coupling, some units have a lower number of connections \emph{(degree)}.
Such a unit is less ``held back'' via the diffusive coupling and can thus get excited more easily.
This effect may in turn tip the scales for certain units, which already have properties beneficial for excitation (e.g., a low \(b_i\)), such that they can become excited from time to time.
These units form the source of a patch of excitation that can start a spiral wave or extreme event.
This allows trajectories to escape from attractor~\Atilde and thus makes it less prevalent when the inhomogeneity of the coupling is increased.
Attractor~\Btilde (traveling waves) also becomes less prevalent with the increase of irregularities in the coupling topology, as they facilitate wave patterns to be non-periodic, which in turn makes them likely to eventually cease by wavefronts canceling out one another or to culminate in an extreme event.
Finally, extreme events (pattern~C) and global oscillations (attractor~\Ctilde) being facilitated by long-range connections is in accordance to our observations from Sec.~\ref{phenomenological}:
Due to long-range connections, excitations can advance more quickly and excite other units than they would in case of only short-range connections, thus causing events.

\emph{Size of the sphere of local influence}
--~Secondly, we investigate the impact of the size of the sphere of local influence on pattern switching by modifying the number~\(m\) of units that constitute the sphere (data not shown).
If we enlarge the sphere's size by up to 50\,\%, we again observe pattern switching with all three patterns.
On the other hand, shrinking the sphere of local influence leads to more frequent self-generated excitations, which results in global oscillations~(\Ctilde).
In line with our above explanation, this can be related to an increasing portion of units with a comparably low degree~\footnote{This increase is due to the strong increase of the coefficient of variation of the degree distribution with decreasing~\(m\) (e.g., the coefficient of variation amounts to \(0.21\) for \(m=8\) and to \(0.08\) for \(m=60\)).}.
Massively enlarging the sphere of local influence, however, results in a completely coupled network, which only exhibits a switching between patterns A and~C, but no wave-like phenomena comparable to pattern~B \cite{Ansmann2013, Karnatak2014}.

\emph{Dimension of the underlying lattice}
--~Lastly, we address the question whether pattern switching can also be observed on one-dimensional lattices (with \(n=100\) and \(m=8\), which corresponds to one row of our two-dimensional lattice, or with \(n=10000\) and \(m=60\)).
On these networks, we still observe patterns A and~C as well as switchings between them.
However, those systems do not exhibit a one-dimensional analogue to pattern~B with a long but finite duration.
This is to be expected insofar as any wave-like phenomena comparable to pattern~B can either only travel a finite distance before extinguishing each other upon collision or will travel unimpededly forever.

\subsection{Parameter inhomogeneity}

We now discuss the estimated relative frequencies of adopted attractors and pattern switchings depending on parameter inhomogeneity (right panel of Fig.~\ref{fig:p_dependency}).
When the units become more inhomogeneous, we observe the following sequence of regimes in parameter space:
1\textsuperscript{0}~global oscillations (attractor~\Ctilde), pattern switching, 1\textsuperscript{1}~global oscillations (attractor~\Ctilde), again pattern switching and finally low-amplitude oscillations (attractor~\Atilde).
The regimes dominated by pattern switching and 1\textsuperscript{1}~global oscillations feature a small fraction of realizations exhibiting traveling waves (attractor~\Btilde).

Our observations demonstrate that control-parameter inhomogeneity suppresses a strongly synchronized dynamics (1\textsuperscript{0} global oscillations).
However, if the inhomogeneity is too high, dynamics involving excitations, namely patterns B and~C as well as attractors \Btilde and~\Ctilde, do not occur.
This can be explained by a larger portion of units being comparably difficult to excite (due to having a high value of~\(b_i\), which governs how strong the inhibitory variable~\(y_i\) reacts to the excitatory variable~\(x_i\)) and thus may suppress the formation of patches of excitation or the spread of excitations.
Thus dynamical behaviors other than low-amplitude oscillations are suppressed and attractor~\Atilde becomes dominant.

The fact that a regime with a dominance of global oscillations separates the regimes with a dominance of pattern switching suggests that the latter occurs in between different mixed-mode-oscillatory windows in parameter space, corresponding to several attractors of type~\Ctilde.
Those windows are ``smeared out'' due to the influence of the system's realizations.
This resembles previous findings on irregular extreme events occurring in chaotic windows in between mixed-mode-oscillatory windows~\cite{Karnatak2014}.
Finally, in contrast to the dependence on the coupling inhomogeneity (see Fig.~\ref{fig:p_dependency}, left), there is no control-parameter inhomogeneity for which most realizations exhibit convergence to attractors of type~\Btilde.
This suggests that, in contrast to inhomogeneities in the coupling topology, parameter inhomogeneities are not relevant for attractor~\Btilde losing or attaining prevalence by the presence or absence, respectively, of the switchings \BtoA and \BtoC.

\vfill

Summarizing this section, we conclude that a two-dimensional coupling topology containing both, short\nobreakdash- and long-range connections, a moderately sized sphere of local influence, and a certain amount of inhomogeneity in control parameters are required for pattern switching between all three patterns to occur in the system.

\section{Recurrences and lifetimes}\label{statistics}

Next we investigate whether regularities can be identified in the lifetimes of patterns and in the pattern sequence.
From 100 trajectories of the same system as in Sec.~\ref{phenomenological} but with different initial conditions, we observe pattern switching in 92~cases.
In 6~cases, the trajectory converges to attractor~\Ctilde (1\textsuperscript{1}~global oscillations); and in 2~cases, it converges to attractor~\Btilde (traveling waves).
This suggests that attractor~\Ctilde has the larger basin of attraction.

\begin{figure*}
	\includegraphics{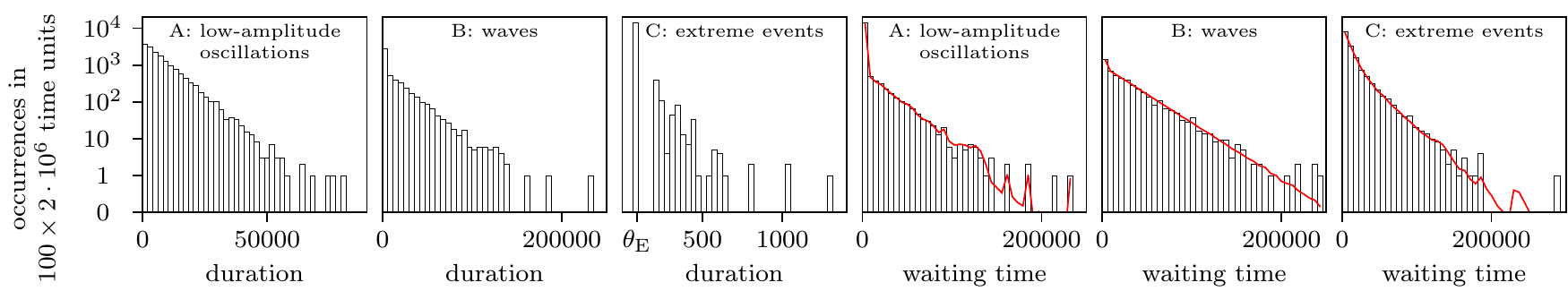}
	\caption{(Black) Histograms of pattern durations and waiting times (end to beginning) between subsequent occurrences of one pattern in \(100 \times 2 \cdot 10^6\) time units.
	For the former, only patterns that began and ended during the observation time were taken into account.
	(Red) The same averaged over 200 simulations of the pattern sequences (of the same length as the originally observed pattern sequences) as second-order Markov chains with durations of patterns randomly assigned from the respective distributions shown in the left half.
	Occurrences of pattern~C that are longer than \(\theta_\text{E}=90\) are due to a few events occurring in short succession.
	They contain at most six events and make up for less than 5\,\% of all occurrences of pattern~C.}
	\label{fig:histograms}
\end{figure*}

The distributions of lifetimes of patterns A and~B (Fig.~\ref{fig:histograms}) are nearly exponential and thus resemble those for a Poisson process, the strongest deviation being a surplus of short occurrences of pattern~B.
Occurrences of pattern~C have a much lower typical lifetime, comprising only one extreme event in the majority of cases and six extreme events at most.
The respective waiting times between occurrences of pattern A and~B are also nearly exponentially distributed (Fig.~\ref{fig:histograms}), which is to be expected given the exponentially distributed lifetimes of patterns B and~A, respectively, and the comparably short lifetimes of pattern~C.

\begin{table}
	\begin{tabular}{c r @{~~\quad} c}
		pattern sequence & \multicolumn{1}{c}{occurrences~~} & probability \\\hline
		B\textrightarrow A\textrightarrow B & 451 & 0.25 \\
		B\textrightarrow A\textrightarrow C & 1369 & 0.75 \\[2pt]
		C\textrightarrow A\textrightarrow B & 3592 & 0.25 \\
		C\textrightarrow A\textrightarrow C & 10534 & 0.75 \\[2pt]
		A\textrightarrow B\textrightarrow A & 1511 & 0.37 \\
		A\textrightarrow B\textrightarrow C & 2545 & 0.63 \\[2pt]
		C\textrightarrow B\textrightarrow A & 314 & 0.38 \\
		C\textrightarrow B\textrightarrow C & 520 & 0.62 \\[2pt]
		A\textrightarrow C\textrightarrow A & 11711 & 0.98 \\
		A\textrightarrow C\textrightarrow B & 249 & 0.02 \\[2pt]
		B\textrightarrow C\textrightarrow A & 2477 & 0.81 \\
		B\textrightarrow C\textrightarrow B & 588 & 0.19
	\end{tabular}
	\caption{Number of occurrences of length-3 subsequences of the pattern sequence in \(100 \times 2 \cdot 10^6\) time units and their respective probabilities given a fixed middle pattern.
	Only for subsequences with pattern~C (event) in the middle, does the last pattern significantly depend on the first one (Fisher's exact test~\cite{Fisher1922}, Bonferroni-corrected, significance level \(0.05\)).
	For longer subsequences (up to length~18), no comparable dependencies were found.
	}
	\label{tab:transitions}
\end{table}

Testing for regularities in the pattern sequence, we find that the sequence can be described by a second-order Markov process without a longer-lasting memory (Table~\ref{tab:transitions}).
We also simulated the pattern sequence as generated by this Markov chain and assigned each occurrence of a pattern a random lifetime from the corresponding distribution.
The resulting distributions of waiting times are in good accordance with the distribution of waiting times observed for our system (see Fig.~\ref{fig:histograms}).

Our findings show that the pattern sequence as well as the termination of individual patterns strongly resemble random processes.
On the other hand, pattern switching as a whole appears to be a transient dynamics, possibly supertransients~\cite{Tel2008} or stable chaos~\cite{Politi2010}.

\begin{figure*}
	\includegraphics{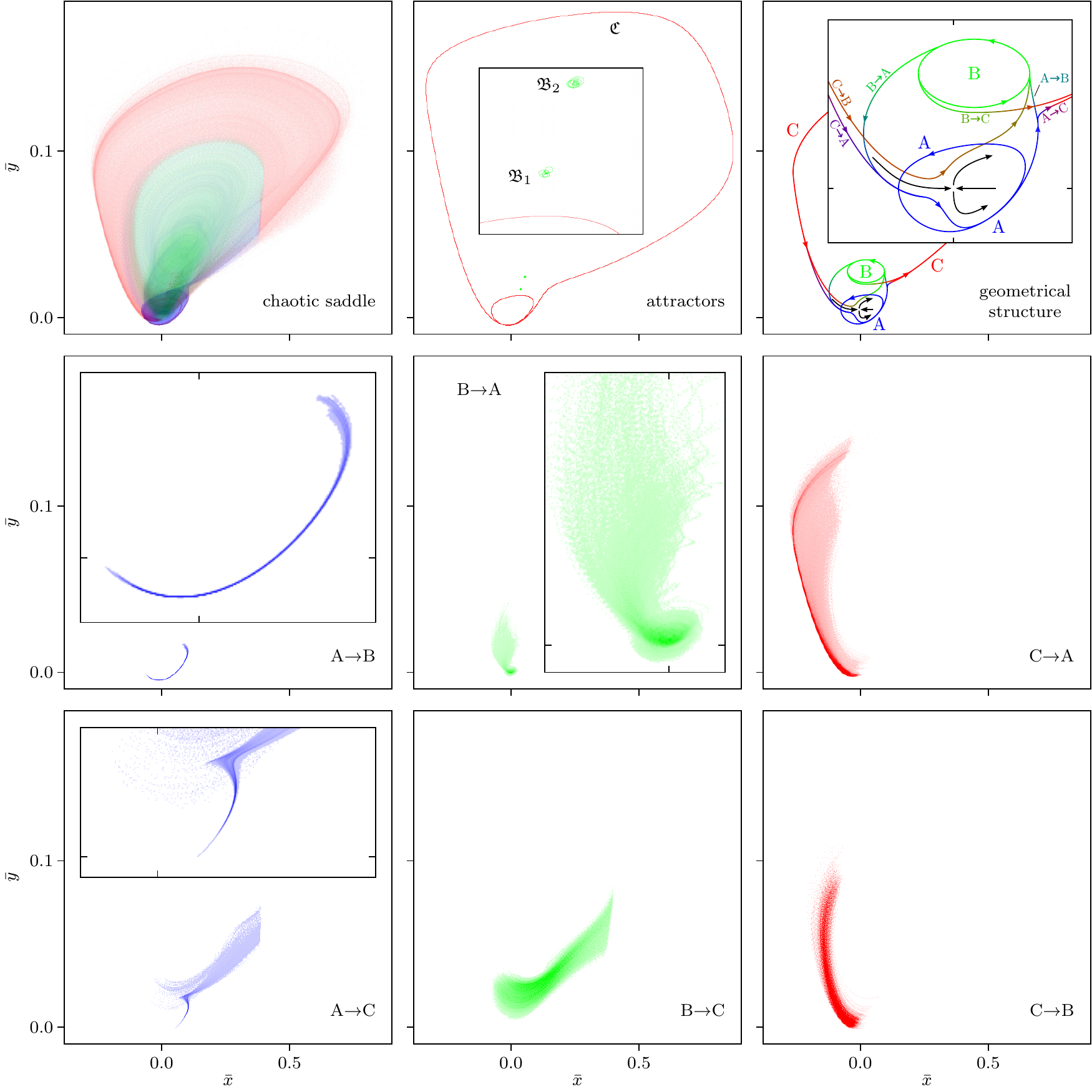}
	\caption{(Top left): Scatterplot of \(\bar{x}\) and~\(\bar{y}\) for \(100\) trajectories (the same as used for Fig.~\ref{fig:histograms} and Tab.~\ref{tab:transitions}) with the point color indicating the respective pattern (blue:~A, green:~B, red:~C), excluding trajectory segments from the attractors.
	(Top middle): Same, only with data from the attractors \Btilde (traveling waves) and \Ctilde (1\textsuperscript{1} global oscillations).
	There are two distinct attractors for traveling waves (\(\mathfrak{B}_1\) and \(\mathfrak{B}_2\)), one manifested as a horizontally traveling wavefront, one manifested as a diagonally traveling wavefront.
	Note that their phase-space projections are almost a point, because \(\bar{x}\) and~\(\bar{y}\) change only little over one period.
	Attractor~\Atilde was not observed for this realization of the coupling topology and control-parameter inhomogeneity.
	(Middle and bottom row): Same with data only from 50~time units before switchings between patterns.
	(Top right): Sketch of stable and unstable manifolds (black) associated with the fix point at the origin and typical trajectories (colors encode patterns as above, trajectories corresponding to switchings are shown in blended colors).
	(All panels): Insets show zooms of the respective plots.
	Trajectories are generally directed counter-clockwise.}
	\label{fig:phasenraum}
\end{figure*}

\section{Manifolds and phase space}\label{channels}

We now discuss pattern switching in our system from a dynamical point of view, analyzing the structure of the phase space and the possible arrangements of stable and unstable manifolds, which would support the observed switching dynamics.
In the following, we demonstrate that a large chaotic saddle containing the three space--time patterns connected by channel-like structures is the backbone of the pattern-switching dynamics.

A chaotic saddle is a complicated, possibly fractal set in phase space on which the dynamics is chaotic, but which has the character of a saddle, i.e., it possesses stable and unstable manifolds (more precisely, foliations) comparable with stable and unstable manifolds of a saddle point.
Since the chaotic saddle is an unstable invariant set, the trajectory will finally escape and converge to an attractor~\cite{Lai2011}.
The emergence and characteristics of a chaotic saddle in transient spatiotemporal chaos have been reported previously at the transition to turbulence in shear flows \cite{Hof2006, Joglekar2015}, reaction-diffusion systems~\cite{Wacker1995}, in the complex Ginzburg-Landau equation~\cite{Houghton2010} and in neuronal networks~\cite{Keplinger2014}.
Most of these examples, however, have in common that the chaotic saddle is characterized by one type of spatiotemporal dynamics.
Recently it has been reported that in transient spatiotemporal chaos the alternation between two different space--time patterns either on the whole spatial domain~\cite{Lafranceschina2015} or on two subdomains~\cite{Haugland2015} can be observed.
By contrast, the chaotic saddle identified here has a more complicated structure since it contains three different space-time patterns and an irregular switching between them mediated by channel-like structures in phase space.
To our knowledge, such a saddle has not been described before and extends the notion of a chaotic saddle in spatially extended systems.
In the following, we unravel the internal structure of this chaotic saddle and with it the mechanism of pattern switching observed before the trajectory converges to one of the attractors of type \Btilde~(traveling waves) or \Ctilde~(global oscillations), respectively.

In Fig.~\ref{fig:phasenraum}, we present a projection of our system's 20000-dimensional dynamics onto a two-dimensional phase-space using the mean values \(\bar{x}\) and \(\bar{y}\) over all oscillators in the network.
The chaotic saddle in our system extends over most of the phase space and contains the three distinct space-time patterns A, B, and~C, which are depicted with different colors in the upper left panel of Fig.~\ref{fig:phasenraum}.
The attractors to which some of the trajectories escape are shown in the upper middle panel of Fig.~\ref{fig:phasenraum}.
In this projection onto a two-dimensional space, it seems that attractor \Ctilde (global oscillations) surrounds the chaotic saddle, which in turn surrounds the attractors of type \Btilde (traveling waves).

To understand how trajectories can switch between the different patterns contained in the chaotic saddle, we have to investigate its internal structure in detail.
To this end, we single out trajectory segments corresponding to time intervals just before switching.
Trajectory segments for the switching \AtoC all pass through a very narrow channel-like structure (Fig.~\ref{fig:phasenraum}, bottom left).
Similar channel-like structures were described before to underlie generation of extreme events \cite{Ansmann2013, Karnatak2014}.
The channel-like structure for \AtoB (Fig.~\ref{fig:phasenraum}, middle left) is at first almost parallel, very close, though not overlapping with the channel-like structure for \AtoC, before it branches off to a different direction.
This is in accordance with our observation that those two switchings are very similar phenomenologically (see Sec.~\ref{phenomenological}) as they both originate from small patches of excitation.

The channel-like structures for the switchings \BtoA and \BtoC appear to be broader in this projection onto a two-dimensional space.
For \BtoA, the trajectory comes very close to the saddle fixed point at the origin \(\kl{0,0}\).
This behavior corresponds to the observation from Sec.~\ref{phenomenological} that no unit shows any excitation at this point.

For the switchings from pattern~C, we find again channel-like structures for \CtoA and a less narrow one for \CtoB (Fig.~\ref{fig:phasenraum}, right middle and bottom).
The similarity of these channels is in accordance with the phenomenological similarity of the switchings, which both result from a large number of refractory units  (see Sec.~\ref{phenomenological}).
For these switchings, the saddle fixed point at the origin and its stable and unstable manifolds appear to play a pronounced role.
Both these channels (\CtoA and \CtoB) come very close to the saddle fixed point at the origin but seemingly on two different sides of its stable manifold:
If the trajectory approaches the saddle fixed point beneath its stable manifold, it turns to pattern~A; if it approaches it above the manifold, it turns to pattern~B.
Note that the saddle fixed point and its stable and unstable manifolds are not part of the chaotic saddle but pertinent for its location in phase space.

Putting all these channel-like structures together, we sketch in the upper right panel of Fig.~\ref{fig:phasenraum}, a comprehensive description of the mechanisms of pattern switching illustrating the pathways of switchings on the chaotic saddle and highlighting the crucial role of the saddle fixed point and its stable manifold which acts as a junction.
Please note that the sketch represents the patterns in an abstract form not reflecting their proper shape.

\section{Discussion}\label{discussion}

We reported on a self-induced switching between more than two distinct space--time patterns---low-amplitude oscillations, nonlinear waves, and extreme events---on a spatially extended excitable system.
This phenomenon is neither caused by a heteroclinic orbit, as the patterns succeed each other randomly, nor by an external signal or noise, as the latter do not exist in our model.
We studied pattern switching from different points of view---phenomenologically, statistically, as well as dynamically and geometrically---, which yielded a coherent picture of the underlying mechanism and crucial ingredients.

Our findings imply that the observed pattern switching is a very long transient on a chaotic saddle that contains all three distinct space--time patterns, connected by channel-like structures mediating the switching.
Such a configuration has, to our knowledge, not yet been demonstrated, and it extends the usual notion of a chaotic saddle in spatially extended systems, in the sense that it provides a mechanism for switchings between more than two distinct space-time patterns.
Transient spatiotemporal chaos studied in other systems governed by a chaotic saddle usually exhibits one type of dynamics \cite{Lai1995, Rempel2007} or at most two alternative patterns as observed in a neuronal network~\cite{Lafranceschina2015} or in a modified complex Ginzburg--Landau equation on two spatial subdomains~\cite{Haugland2015}.

The aforementioned configuration may provide a general mechanism for pattern switching in system classes beyond spatially extended excitable systems.
For the latter, our investigations revealed two system properties that are crucial to obtain pattern switching:
\begin{itemize}
	\item The units are inhomogeneous, but are all in the oscillatory regime and, if coupled, they are capable of self-generating localized excitations.
	\item The coupling topology is dominated by connections that are short-ranged with respect to an at least two-dimensional geometry, but also contains a certain amount of long-range connections.
\end{itemize}
The former capability, i.e., self-generating excitations, was related to an interior crisis in other systems \cite{Ansmann2013, Karnatak2014}.
However, these systems do not exhibit pattern switching between more than two distinct space--time patterns, and we also found no other straightforward simplification of our system that does exhibit such a switching.

In the space of parameters that control the dynamics of the individual units and the coupling topology of our system, pattern switching covers a rather large region that is surrounded by regimes where the trajectories converge to attractors.
Changing the number of long-range connections or the control-parameter inhomogeneity (see Sec.~\ref{parameters}) alters the arrangement of attractors and saddles in phase space, particularly the location of their stable and unstable manifolds.
This in turn leads to the opening of channel-like structures through which pattern switching is enabled.

The switching dynamics can further be elucidated by borrowing concepts and properties for leaking chaotic systems~\cite{Altmann2013}.
Such systems possess a transient chaotic dynamics, whose trajectories eventually escape through a leak, after an exponentially distributed life time.
In our system, each pattern is analogous to a leaking chaotic system (in particular, the life times of patterns A and~B are exponentially distributed) and the channel-like structures correspond to leaks.
While for leaking chaotic systems, the sites and locations of leaks are explicitly prescribed, the channel-like structures within the chaotic saddle of our system appear due to changes of the arrangement and location of attractors, saddles, and manifolds in phase space when control parameters are varied.
The size of the channel-like structures can be related to the rate of emergence of extreme events in the same way as mean escape times in leaking chaotic systems scale with the size of the leak~\cite{Karnatak2014}.

The existence of this chaotic saddle with embedded channels, and thus the phenomenon of switching between multiple patterns, is robust to moderate changes of control parameters.
Moreover, the phenomenon is not tied to a specific system size, as it occurred for systems with about the half, double and quadruple number of units (data not shown).
Nevertheless, investigating the scaling behavior in the characteristics of this phenomenon is an open challenge due to computational constraints.
Such an investigation could reveal the dependence of the excessive transient length on various parameters, such as the number of nonlocal couplings~\cite{Yonker2006} and the system size \cite{Kaneko1990, Strain1998, Wackerbauer2003, Zumdieck2004, Houghton2010}.
The latter would allow to judge whether the chaotic saddle turns into an attractor with the switching becoming permanent for an infinitely large system \cite{Tel2008, Politi2010}.
An improved characterization of the chaotic saddle could be obtained from computing escape times from the chaotic saddle~\cite{Lai2011} and their dependence on the particular pattern from which the escape happens.

Self-induced pattern switching may also be observed for other dynamics of individual units, other types of coupling, or other types of networks.
With respect to the latter, we expect other coupling topologies, such as hierarchical or modular ones~\cite{Arenas2008}, which feature different kinds of connections, to allow for similar phenomena.
As for the individual units, we chose them to be in an oscillatory regime when uncoupled.
This suggests, that pattern switching might not be restricted to excitable units but could be possible also in other systems where each unit exhibits relaxation oscillations.
However, such systems should bear two additional properties:
The coupling should be such that it suppresses high-amplitude oscillations in order to facilitate low-amplitude oscillations, and the system should be capable of self-generating excitations.

\section{Conclusions}\label{conclusion}

Self-induced switchings between multiple different space--time patterns, including extreme events, can be facilitated by a chaotic saddle that contains these patterns and channel-like structures between them, allowing for the switching.
Such a mechanism offers an explanation for the variety of observable dynamical behaviors and switchings between them in natural systems.
For spatially extended excitable systems, we show the self-generation of extreme events from different space--time patterns and their ending in possibly other space--time patterns.
Such a behavior is known for example for the heart and the brain --~spatially extended excitable systems with properties we found crucial for pattern switching.
Indeed, the heart and the brain exhibit switchings between patterns associated with normal functioning~\cite{Rabinovich2011} as well as between normal and pathological, extreme behaviors, such as epileptic seizures~\cite{Engel2007, Lytton2008}, migraine attacks~\cite{Pietrobon2003, Dahlem2004b}, or atrial or ventricular fibrillations~\cite{Sankaranarayanan2008, Clayton2011, Qu2014}.

\begin{acknowledgments}
The authors would like to thank C.~Grebogi, R.~Karnatak, E.~Meron, A.~Politi, D.~Postnov, T.~T\'el, H.~Uecker for interesting discussions.
We are indebted to the unknown reviewers for their helpful comments.
We are grateful to S.~Bialonski, A.~Saha, and K.~Stahn for critical comments on earlier versions of the manuscript.
This work was supported by the Volkswagen Foundation (Grant~Nos. 85388, 85392, 88459, and 88463 within the call Extreme Events: Modeling, Analysis and Prediction).
\end{acknowledgments}

\appendix

\section{Largest Lyapunov exponents}\label{lyap_avg}\label{lyap_local}
To estimate local and time-averaged largest Lyapunov exponents, we employ the established approach of evolving a tangent vector parallel to the actual dynamics~\cite{Benettin1980}.
We rescale the tangent vector every \(\tau\) time units and denote its length before rescaling at a time~\(t\) as \(\alpha\kl{t}\).
As an estimate of the largest local Lyapunov exponent, we use \(\lambda\kl{t} \defi \tfrac{1}{\tau} \log\kl{\alpha\kl{t+\tau}}\).
As an estimate of the largest time-averaged Lyapunov exponent, we average \(\lambda\kl{t}\) over an observation time of at least \(T=200000\) time units.
In both cases, we used \(\tau=1.0\).

To determine the confidence of the sign of the largest time-averaged Lyapunov exponent, we check whether the mean of \(\klg{\lambda\kl{0}, \ldots, \lambda\kl{T}}\) significantly deviates from~\(0\) (Student's one-sample \(t\)-test).
If yes, we conclude the Lyapunov exponent to have the corresponding sign, otherwise we evaluate it to be zero.
In all investigated cases in which we conclude a signed Lyapunov exponent, the error probability is \(0\) within the limits of numerical accuracy; in all other cases, it is larger than~\(0.9\).

\end{document}